\documentstyle[prb,aps,epsf]{revtex}
\begin{document}
\title{Vortex lattice and matching fields for a long 
superconducting wire} \draft
\author{Pablo A. Venegas\cite{santacruz} \\
Physics Department, University of California \\
Santa Cruz, CA 95064}
\author{Edson Sardella}
\address{Departamento de F\'{\i}sica, Universidade Estadual Paulista \\
Caixa Postal 473, 17033-360, Bauru-SP, Brazil}
\date{\today}
\maketitle
\begin{abstract}

  We investigate the flux penetration patterns and matching fields of a long
cylindrical wire of circular cross section in the presence of an external magnetic
field. For this study we write the London theory for a long cylinder both for the
mixed and Meissner state, with boundary conditions appropriate for this geometry. 
Using the Monte Carlo Simulated Annealing method, the free energy of the mixed
state is minimized with respect to the vortex position and we obtain the ground
state of the vortex lattice for $N=3$ up to 18 vortices. The free energy of the
Meissner and mixed state provides expressions for the matching
fields. We find that, as in the case of samples of different geometry, the finite
size effect provokes a delay on the vortex penetration and a vortex accumulation in
the center of the sample. The vortex patterns obtained are in good agreement with
experimental results. \end{abstract}


\section{INTRODUCTION}

 Modern technology has made possible to fabricate superconducting samples of small
size, like films of thickness less than the London penetration length as well as
superconducting wires of radii of order of this length. This has arisen the
interest to study again the geometrical or size effects in superconductors.  The
size problem was already studied a long time ago \cite{guyon,brandt1} but recently
it has been reconsidered and some studies of transport currents, magnetization,
magnetic moment, reversibility lines, and flux penetration in finite size samples
have been made
\cite{brandt1,brandt2,brandt3,shuster,zeldov1,zeldov2,exp1,exp2,kaykovich}. 
 In particular, the penetration of magnetic flux in finite size superconductors has
attracted the attention of physicists in the last years due to the fact that in
finite samples the geometrical effects can produce important modifications in the
critical state and in the resulting vortex distribution. As an example we can cite
the results of Zeldov et al. \cite{zeldov2} in thin superconducting strips which
have been found a delay in the penetration of the vortex lines and a vortex
accumulation at the center of the sample due to the geometrical barrier effect, but
other studies at this respect has been made \cite{brandt1}. In this way, our main
interest in this work is to study specifically the size effects in the vortex
lattice and in the matching fields (the minimum field for a new vortex 
penetration) for an infinite superconducting wire of circular
cross section. 

We use the London theory to study the formation of vortex lattice in the mixed
state of a long superconducting cylinder.  London theory is valid in the limit of
low induction (fields well bellow the upper critical field $H_{c2}$), where most of
the experiments can be performed to observe vortex lattice.  This theory fails in
the limit of small length scale. This breakdown of the theory is originated in the
fact that the finite size of the vortex cores is neglected. Thus, London theory is
not suitable to treat the self-energy of a single vortex line.  However, what
really matters to determine the shape of the ground state of the vortex lattice is
the interaction energy between vortices on different sites. For an infinite sample,
vortices interact one with each other via a two-body potential.  Nevertheless, if
the surface is taken into account, additional terms which describe the interaction
of the vortex with the surface come in. In this work we develop the theory to
describe such interactions in a long cylinder.  In addition, we use London theory
to determine the free energy of the Meissner state of a long superconducting wire. 
 For the determination of the ground state of the vortex lattice we use
the Monte Carlo Simulated Annealing minimization method. For the first
time, we obtain the ground state lattice pattern starting from an
arbitrary configuration.  This procedure is different than the usually
used where the free energy of the vortex lattice for several predetermined
configurations is calculated and pick the lowest one.

The paper is outlined as follows. In Section \ref{sec2} we determine the magnetic
field of an arbitrary distribution of vortices in the mixed state and its
corresponding induction (spatial average of the local magnetic field). In addition,
we find both the London and Gibbs free energy. In Section \ref{sec3} we repeat this
calculation to the Meissner state. In Section \ref{sec4} and \ref{sec5} we analyze
the  matching fields and the vortex lattice patterns. 

\section{MIXED STATE}\label{sec2} In what follows we develop a theory for
the mixed state of a long superconducting wire.  Our starting point is the
London equation. This equation is obtained from the second Ginzburg-Landau
equation by assuming that the superconducting order parameter is a
constant throughout the whole space, that is, it neglects variations of
the order parameter inside the vortex-core.  The London approximation is
valid provided that the Ginzburg-Landau parameter $\kappa=\lambda/\xi\gg
1$; here $\xi$ is the coherence length, and $\lambda$ the London
penetration length. In addition, the vortices, whose size is of the order
of $\lambda$ , may overlap but not the vortex-cores. In cylindrical polar
coordinates $(r,\phi)$ the London equation for the local magnetic field
${\bf h}=h{\bf z}$ is given by

\begin{equation}
-\lambda^2\left ( \frac{\partial^2h}{\partial r^2}+
\frac{1}{r}\frac{\partial h}{\partial r}+
\frac{1}{r^2}\frac{\partial^2h}{\partial \phi^2}\right ) +h=
\Phi_0\,\sum_i\,\delta({\bf r}-{\bf r}_i)\;,
\label{londonequation}
\end{equation}
where $\Phi_0$ is the quantum flux, ${\bf r}_i$, is the position
of the $i$-th vortex inside the cylinder, and $\delta({\bf r})=
\delta(x)\delta(y)$ is the two dimensional delta function. Here we
are assuming that the vortices are straight lines. Therefore, a 3D
problem is reduced to a 2D one. We will solve this equation
subject to the following boundary conditions

\begin{eqnarray}
h(a,\phi) & = & H \nonumber \\
\left( \frac{\partial h}{\partial \phi}\right )_{r=a} & = & 0\;,
\label{hboundarycondition}
\end{eqnarray}
where $a$ is the radius of the cylinder. The first condition
assures that outside the sample the field is uniform and is
precisely the external field $H$; the second one states that the
perpendicular component of the current vanishes at the boundary
$r=a$, that is, the Cooper pairs cannot jump out of the
sample.

To solve (\ref{londonequation}) we use the Green's function method. 
Assuming for the Green function the boundary conditions 
$G(a,\phi,r^{\prime},\phi^{\prime})  =  0 $, 
$G(r,\phi,r^{\prime},\phi^{\prime})$ continuous at $r=r^{\prime}$,
 $\partial G(r,\phi,r^{\prime},\phi^{\prime})/\partial r$
 discontinuous  at $r=r^{\prime}$ and assuming
that both $G$ and $h$ are periodic in $\phi$, one obtains

\begin{equation}
h(r^{\prime},\phi^{\prime})=\Phi_0\,\sum_i\,
G(r_i,\phi_i,r^{\prime},\phi^{\prime})-Ha\lambda^2\,
\int_0^{2\pi}\,d\phi\,\left ( \frac{\partial G}
{\partial r} \right )_{r=a}\;.
\label{field}
\end{equation}

To proceed we need to solve the equation for the Green's
function. The method we use to find this function is outlined 
in Ref. [\onlinecite{jackson}] except by the fact 
that there the Green's function is associated with the Poisson 
equation and boundary conditions are taken at infinity. One has

\begin{equation}
G(r,\phi,r^{\prime},\phi^{\prime})=\frac{1}{2\pi\lambda^ 2}\left [
K_0(|{\bf r}-{\bf r}^{\prime}|/\lambda)-
\sigma(r,\phi,r^{\prime},\phi^{\prime})
\right ]\;,
\label{gfinal}
\end{equation}
where

\begin{equation}
\sigma(r,\phi,r^{\prime},\phi^{\prime})=
\sum_{m=-\infty}^{+\infty}\,\cos[m(\phi-\phi^{\prime})]
\frac{K_m(a/\lambda)}{I_m(a/\lambda)}
I_m(r/\lambda)I_m(r^{\prime}/\lambda)\;,
\label{sigma}
\end{equation}
where $I_m$ and $K_m$ are the modified Bessel functions.

We are now in a position to find the local magnetic field. Substituting
(\ref{gfinal}) into (\ref{field}) and using the identity 
$I_m(x)K_m^{\prime}(x)-I_m^{\prime}(x)K_m(x)=-\frac{1}{x}$ (where the primes on $I$
and $K$ stands for the first derivative with respect to $x$) to
develop the second term for the field expression one
obtains

\begin{equation}
h(r,\phi)=\frac{\Phi_0}{2\pi\lambda^2}\,\sum_i\,
[K_0(|{\bf r}-{\bf r}_i|/\lambda)-
\sigma(r,\phi,r_i,\phi_i)]+H\frac{I_0(r/\lambda)}
{I_0(a/\lambda)}\;.
\label{hfinal}
\end{equation}

The London (Helmholtz in the thermodynamic context) free energy contains
basically two contributions. One is the energy stored in the field and the
other one is the kinetic energy of the supercurrents. The London
free energy per unit length is

\begin{eqnarray}
\frac{F}{L} & = & \frac{1}{8\pi}\,\int_0^a\,\int_0^{2\pi}\,dr\,d\phi\,r
\left [ h^2+\lambda^2\left ( \frac{\partial h}{\partial r}\right )^2
+\frac{\lambda^2}{r^2}
\left ( \frac{\partial h}{\partial \phi}\right )^2
\right ] \nonumber \\
& = & \frac{\Phi_0}{8\pi}\,\sum_i\,h(r_i,\phi_i)+
\frac{Ha\lambda^2}{8\pi}\,\int_0^{2\pi}\,d\phi\,
\left ( \frac{\partial h}{\partial r}\right )_{r=a}\;,
\label{lfreeenergy}
\end{eqnarray}
where on going from the first to the second line we have used the
London equation (\ref{londonequation}) and the
boundary conditions (\ref{hboundarycondition}) and the
periodicity of the field. Here $L$ is length of the system.

Now, the London free energy can be evaluated by introducing (\ref{hfinal})
into (\ref{lfreeenergy}). This yields,

\begin{equation}
\frac{F}{L}=\left (\frac{\Phi_0}{4\pi\lambda}\right )^2\left \{\,\sum_{i,j}\,
[K_0(|{\bf r}_i-{\bf r}_j|/\lambda)-\sigma(r_i,\phi_i,r_j,\phi_j)]+
\left ( \frac{\tilde{H}}{2} \right )^2\frac{a}{\lambda}
\frac{I_1(a/\lambda)}{I_0(a/\lambda)}
\right \}\;,
\label{lfinal}
\end{equation}
where $\tilde{H}=H/(\Phi_0/4\pi\lambda^2)$.

To obtain the equilibrium configuration of the vortex lattice the
Helmholtz free energy is not convenient because the
calculations involve a fixed magnetic field $H$. Therefore,
it is necessary to perform a Legendre transformation to
obtain the Gibbs free energy. The
Gibbs free energy (in units of volume) is given by,

\begin{equation}
{\cal G}={\cal F}-\frac{BH}{4\pi}\;,\label{gibbs}
\end{equation}
where ${\cal F}=F/AL$, with $A=\pi a^2$, and $B$ is the induction which
is the spatial average of the local magnetic field,

\begin{equation}
B=\frac{1}{A}\,\int\,d^2r\,h\;.\label{induction}
\end{equation}

The evaluation of this integral is tedious but straightforward. We obtain

\begin{equation}
B=\frac{N\Phi_0}{A}+\frac{2\pi a\lambda H}{A}
\frac{I_1(a/\lambda)}{I_0(a/\lambda)}
-\frac{\Phi_0}{A}\frac{1}{I_0(a/\lambda)}\,\sum_{i}\,I_0(r_i/\lambda)\;.
\label{benediction}
\end{equation}

Finally we obtain for the Gibbs free energy

\begin{eqnarray}
{\cal G}& = & \left (\frac{\Phi_0}{4\pi\lambda^2}\right )^2
\frac{\lambda^2}{A}\left [ N\ln\kappa +\,\sum_{i\neq j}\,
K_0(|{\bf r}_i-{\bf r}_j|/\lambda) -  \,\sum_{i,j}\,\sigma(r_i,\phi_i,r_j,\phi_j) 
\right . \nonumber \\
& & \left . + \tilde{H} \,\sum_{i}(\,\frac{I_0(r_i/\lambda)}{ 
I_0(a/\lambda)} - 1\,) - \left ( \frac{\tilde{H}}{2}\right )^2
\frac{a}{\lambda} \frac{I_1(a/\lambda)}{I_0(a/\lambda)}
\right ] \;.
\label{fgibbsfreeenergy}
\end{eqnarray}

Notice that the first term is the vortex self-energy and the second one
describes the repulsive interaction between the vortices (the bulk term). The third
term describes the attractive interaction between the vortices and image vortices
located outside the sample. The effect of this interaction is to push the vortices
close to the surface.  The argument of the fourth term represent the flux
$\Phi/\Phi_0 = (1-I_0(r)/I_0(a))$. The first term on this argument represents the
repulsive interaction between a vortex and the magnetic field that penetrates the
sample surface and it push the vortices to the center of the sample. The second
term in this argument represent the vortex magnetic energy.  The fifth term is the
Meissner state energy (see next Section).  The competition between the
vortex-vortex image interaction and the interaction between the vortex and the
surface field represents an energy barrier that the vortex has to overcome to be
able to enter the sample. When the external field is bellow the matching value, the
vortex-vortex image  interaction is more important and a new vortex is not 
able to
enter.  When the field is increased up to or above the critical value, the vortex
can overcome the surface energy barrier and enters into the sample.  On the other
hand, we want to point out that the bulk interaction is invariant under any
translation but this symmetry is no longer valid for the whole free energy because
of the presence of the various interactions. However the system is still invariant
under any rotation because it depends only on the angle difference. 

\section{MEISSNER STATE}\label{sec3}
In the Meissner state, although we
have no penetration of vortex lines we have penetration of magnetic field
near the surface. In a semi-infinite superconductor, for instance, the
external field penetrates exponentially over a distance $\lambda$. For a
superconductor with cylindrical geometry, the field inside the sample is
given by the London equation,

\begin{equation}
-\lambda^2\left ( \frac{\partial^2h}{\partial r^2}+
\frac{1}{r}\frac{\partial h}{\partial r}\right ) +h=0.
\label{lmequation}
\end{equation}

The boundary condition is

\begin{equation}
h(a)=H\;.\label{bmeissner}
\end{equation}

The solution for this equation with the appropriate boundary conditions is

\begin{equation}
h(r)=H\frac{I_0(r/\lambda)}{I_0(a/\lambda)}\;.\label{meissnerfield}
\end{equation}

Using eq. (\ref{lfreeenergy}) the London free energy per unit length can be 
written as

\begin{eqnarray}
\frac{F}{L} & = & \left (\frac{\Phi_0}{4\pi\lambda}\right )^2
\left ( \frac{\tilde{H}}{2} 
\right )^2\frac{a}{\lambda}\frac{I_1(a/\lambda)}{I_0(a/\lambda)}\;,
\label{lmfreeenergy}
\end{eqnarray}
which is the second term of (\ref{lfinal}). Notice that if $a \gg
\lambda$, near the surface the field has an exponential behavior $h \sim
H(a/r)e^{(r-a)/\lambda}$ like in a semi-infinite superconductor.

Using equation (\ref{induction})  we find for the induction

\begin{equation}
B=\frac{2\pi a\lambda H}{A}\frac{I_1(a/\lambda)}{I_0(a/\lambda)}\;,
\label{bmeisser}
\end{equation}
which is the second term of (\ref{benediction}).

The Gibbs free energy can be calculated by substituting
(\ref{lmfreeenergy}) and (\ref{bmeisser}) into (\ref{gibbs}). One has

\begin{equation}
{\cal G}=-\left (\frac{\Phi_0}{4\pi\lambda^2}\right )^2
\frac{\lambda^2}{A}\left [
\left ( \frac{\tilde{H}}{2} \right )^2\frac{a}{\lambda}
\frac{I_1(a/\lambda)}{I_0(a/\lambda)}\right ] \;,
\end{equation}
which is the forth term of (\ref{fgibbsfreeenergy}).
The previous results comprises the framework for the discussion of the
superconducting properties of a long cylinder. In the next Sections we
examine the matching fields and the vortex lattice patterns.

\section{MATCHING FIELDS}\label{sec4}

The lower critical field $H_{c1}$, defined as the lowest external field
enough to have penetration of at least one vortex line,
is determined assuming that at the phase transition from the Meissner
state to the mixed state, the Gibbs free energy per unit length, i.e.,
${\cal G}A$, has the same value. In what follows, $S$ stands for the mixed
state and $M$ for the Meissner state. One has,

\begin{equation}
A{\cal F}_M-\frac{AB_MH_{c1}}{4\pi}=A{\cal F}_S-\frac{AB_SH_{c1}}{4\pi}\;,
\label{transition}
\end{equation}

where $(A{\cal F}_M,AB_M)$ are given by
(\ref{lmfreeenergy},\ref{bmeisser}) and $(A{\cal F}_S,AB_S)$ by
(\ref{lfinal},\ref{benediction}) respectively.  Introducing these
equations into (\ref{transition}) we find for the lower critical field

\begin{equation}
H_{c1}=\frac{\Phi_0}{4\pi\lambda^2}\left [
\frac{\ln\kappa-\frac{K_0(a/\lambda)}{I_0(a/\lambda)}}
{1-\frac{1}{I_0(a/\lambda)}}\right ] \;.\label{hc1}
\end{equation}

\begin{figure}
\epsfxsize=0.9\columnwidth
\epsfysize=1.1\columnwidth
\centerline{\epsffile{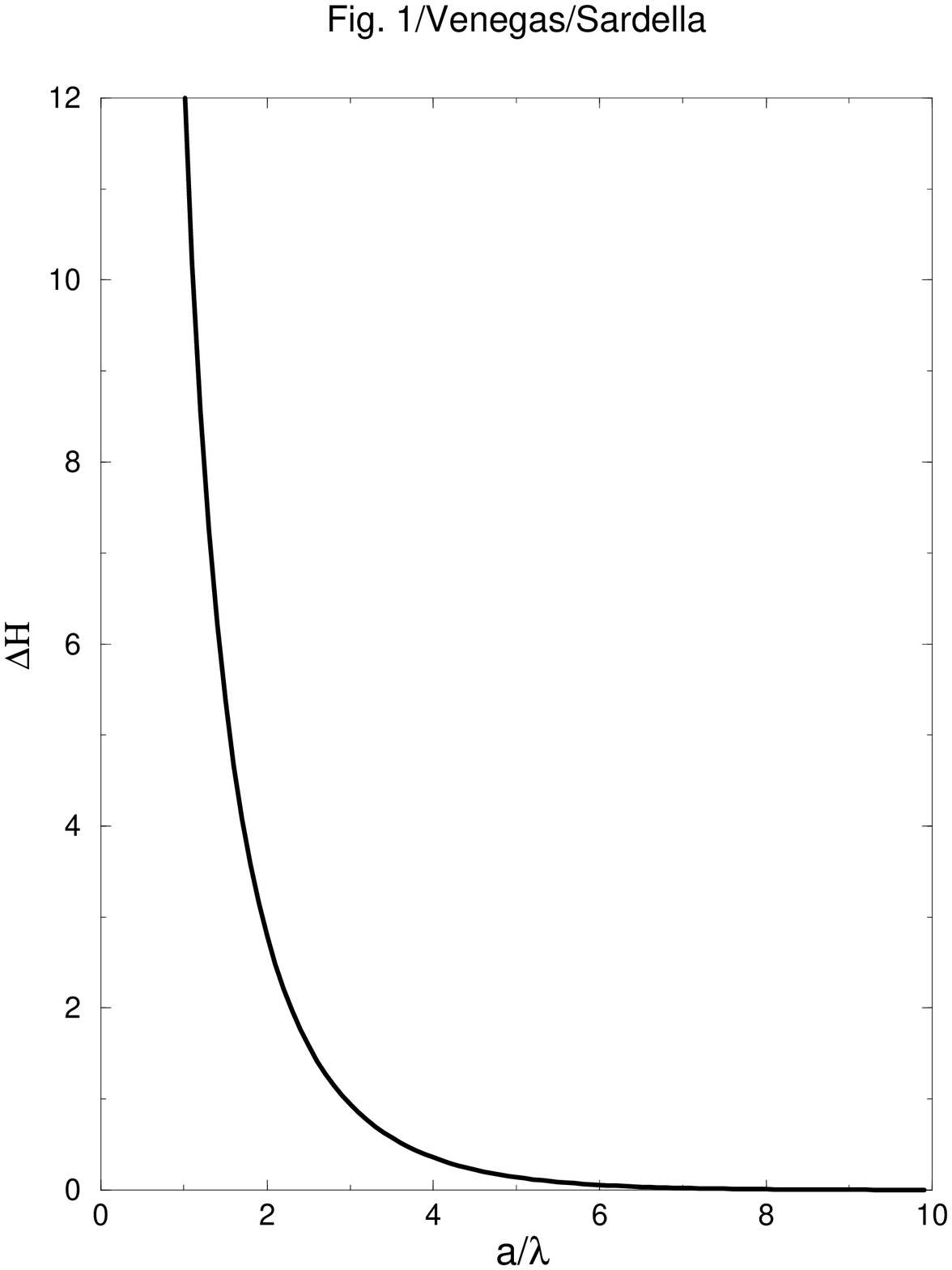}}
\vspace{0.1cm} \caption{The lower critical field difference $\Delta
H=H_{c1}-H_{c1}^{\infty}$ in units of $(\Phi_0/4\pi\lambda^2)$ as a function
of $a/\lambda$ for $\kappa=40$.}
\label{fig1}
\end{figure}

Here we have taken the center of the cylinder as the equilibrium position of a
single vortex line in (\ref{lfinal}) and (\ref{benediction}). In this case, only
the $m=0$ term survives in the sum for $\sigma$ (cf. equation (\ref{sigma})). By
taking the limit of $a \rightarrow \infty$ in (\ref{hc1}) we recover the well known
result $H_{c1}^{\infty}=(\Phi_0/4\pi\lambda^2)\ln\kappa$. In Fig. \ref{fig1} we
plot the difference $\Delta H=H_{c1}-H_{c1}^{\infty}$ in units of
$(\Phi_0/4\pi\lambda^2)$ which shows that the smaller the values of $a$ the larger
the value of the lower critical field. In this way, this result shows clearly that
the size effect provokes a delay in the first flux penetration as found in
experiments on other finite systems of different geometry \cite{brandt1,zeldov1}. 

\begin{figure}
\epsfxsize=0.9\columnwidth
\epsfysize=1.1\columnwidth
\centerline{\epsffile{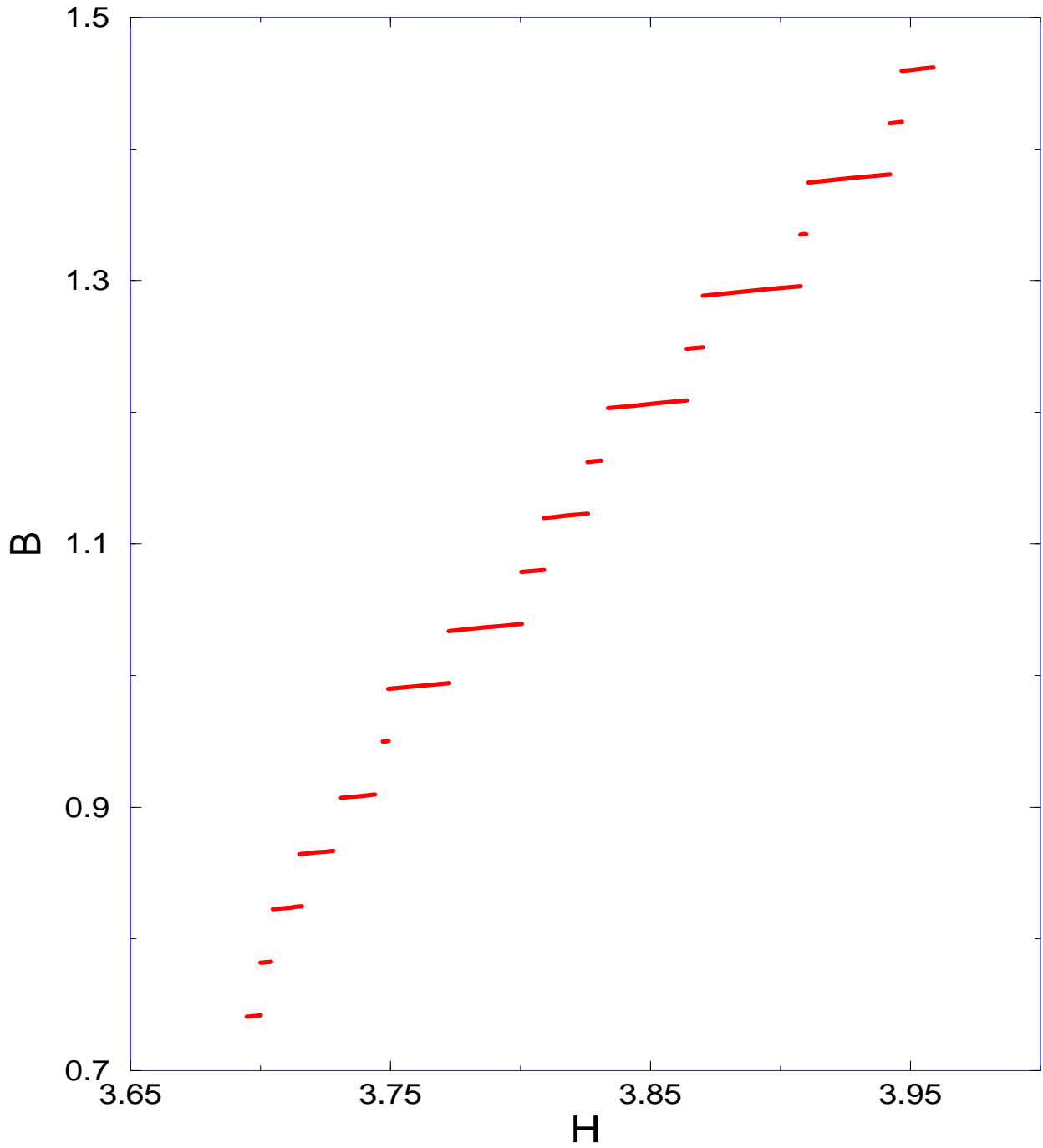}}
\vspace{0.1cm} \caption{Induction ($B$) for $N=1$ to $18$ vortices as a 
function of
the external magnetic field ($ H$) in units of $(\Phi_0/4\pi\lambda^2)$. 
For each $N$ vortices configuration, the induction is calculated in the
 [$H_{sN}$, $H_{sN+1}$] field range. Here we have used $a/\lambda=10$, 
$\lambda=200nm$ and $\kappa=40$.}
\label{fig2}
\end{figure}

As a consequence of the energy barrier generated by the finite size of the sample
we have a delay not only for the first vortex line penetration but for the
subsequent lines too. In this way, at low fields (near the lower critical field),
we have a well defined critical field for each new penetration (matching field).
Different from the bulk case, where the induction increases continuously with
the external field, here the induction increases by steps as showed in Fig.
\ref{fig2}. A new vortex enter in the sample only when the energy is enough to
overcome the surface energy barrier.  As we go to higher magnetic fields the
induction approaches the bulk case. 

The matching fields for each configuration of vortices
($\tilde{H}_{sN}=H_{sN}/(\Phi_0/4\pi\lambda^2)$, $N = 1,..,18$; with
$\tilde{H}_{s1} \equiv \tilde{H}_{c1}$) are calculated using the same procedure
used to obtain $\tilde{H}_{c1}$, i.e.  equating the Gibbs free energy of the
configurations with $N$ and $N+1$ vortices, $G_N \,=\, G_{N+1}$. In this way we
obtain a transcendental equation in $\tilde{H}$ (because the radii implicitly
depends on $\tilde{H}$) which can be solved iteratively. However, each time the
free energy is calculated for a given magnetic field, this free energy must be
calculated with the vortices in the equilibrium position. Thus, for each iteration
the free energy must be minimized. The iterative work was performed using the
Secant method and the minimization using Monte-Carlo Simulated Annealing method
\cite{numrec,goffe}. For $a/\lambda = 10$ the values obtained for $\tilde{H}_{sN}$
can be seen in Table \ref{tab1}.

\section{VORTEX LATTICES}\label{sec5} The usual procedure to find the
ground state of the vortex lattice is to assume some particular geometry
and then evaluate the Gibbs free energy. The configuration corresponding
to the lowest value is the one supposedly the most stable vortex lattice.
Other authors \cite{buzdin,brongersma} have been used the method of
images to determine de vortex configuration. In the present work we follow
a different procedure. Using Monte Carlo Simulated Annealing method
\cite{numrec,goffe}, we start from an initial configuration chosen
randomly and we let the vortex lattice evolve towards the global minimum.
The energy minimization is made using different initial configurations and
different seeds for the random number generator. In this way we obtain
different annealing schedules and assures that the system goes to the
global minimum. This procedure has been usually avoided by many authors
because the computational time strongly increases with the number of
vortices. However, with the advent of very fast computers this is no
longer a major problem. 

As the minimization method may require a large number
of evaluations of the Gibbs free energy, we must find
some effective manner of calculating
$\sigma(r,\phi,r^{\prime},\phi^{\prime})$ (cf. equation (\ref{sigma})),
which involves a sum of
infinite terms. We will consider $a\gg \lambda$, but still finite.
Within this approximation
$I_m(a/\lambda)\approx \frac{1}{\sqrt{2\pi a/\lambda}}e^{a/\lambda}$, and
$K_m(a/\lambda)\approx \sqrt{\frac{\pi}{2a/\lambda}}e^{-a/\lambda}$.
One has,

\begin{equation}
\sigma(r,\phi,r^{\prime},\phi^{\prime})=
\pi e^{-2a/\lambda}I_0(|{\bf r}+{\bf r}^{\prime}|/\lambda)\;,
\end{equation}
where we have used the following identity\cite{prudinokov}

\begin{equation}
\sum_{m=-\infty}^{+\infty}\,\cos(m\phi)I_m(x)I_m(y)=
I_0(\sqrt{x^2+y^2+2xy\cos\phi})\;.
\label{identity2}
\end{equation}

The Gibbs free energy can be simplified to

\begin{eqnarray}
{\cal G} & = & \left (\frac{\Phi_0}{4\pi\lambda^2}\right )^2
\frac{\lambda^2}{A}\left [ N\ln\kappa + \,\sum_{i \neq j}\,
K_0(|{\bf r}_i-{\bf r}_j|/\lambda)-
\pi e^{-2a/\lambda}\,\sum_{i, j}\,I_0(|{\bf r}_i-\overline{\bf r}_j|/\lambda)
\right . \nonumber \\
& & \left . +  \tilde{H}\,\sum_i\,(\frac{I_0(r_i/\lambda)}{I_0(a/\lambda)} - 1)
-\left ( \frac{\tilde{H}}{2} \right )^2\frac{a}{\lambda}
\frac{I_1(a/\lambda)}{I_0(a/\lambda)}
\right ] \;,
\end{eqnarray}
where $\overline{\bf r}_i=-{\bf r}_i$.

\begin{figure}
\epsfxsize=1.1\columnwidth
\epsfysize=1.5\columnwidth
\centerline{\epsffile{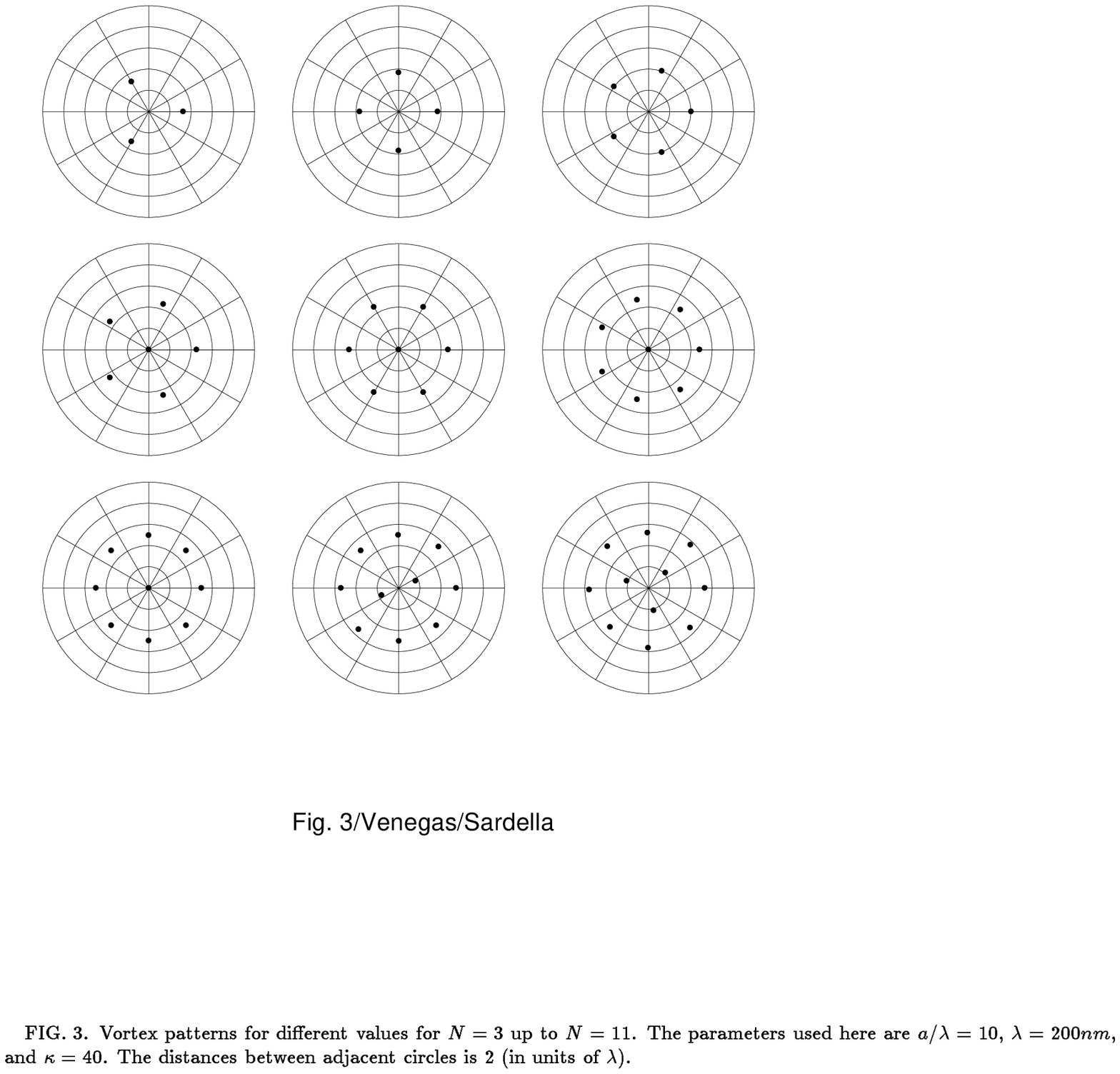}}
\label{fig3}
\end{figure}

To analyze the vortex patterns for each number of vortices we choose
arbitrarily the magnetic field in the middle of the interval between
$\tilde{H}_{sN}$ and $\tilde{H}_{sN+1}$. The minimization of the free
energy was performed again using the Simulated Annealing method.  As we
have observed in Section \ref{sec2}, the system is invariant under any
rotation. Therefore, we can fix one of the vortices along the $x$ axis so
that the minimization procedure will involve $2N-1$ variables. We have
done this for $N=3$ up to $N=18$. The vortex patterns for $N=3$ up to
$N=11$ are illustrated in Fig. \ref{fig3} and the complete numerical data
for $N=3$ to $18$ is shown in Table \ref{tab1}. As it can be seen from the
figure and table, the vortices arrange themselves in quite simple
geometries. For example, for $N=3$ an equilateral triangle; for $N=4$ a
square; for $N=5$ a pentagon; for $N=6$ a pentagon with a vortex at the
center; for $N=7$ a hexagon with a vortex at the center; and for $N=18$ a
vortex at the center, a first ring of six vortices forming a hexagon and a
second ring of eleven almost equally spaced vortices. The results show a
clear tendency of the inner vortices to form a hexagonal lattice, however,
the external not.  For $N=3$ to $9$ the radii of all the vortices in the
ring coincide, however, for $N>9$ the vortices in the outer ring show
small radii fluctuations (no more than $7\%$).  For example, as we can see
on Fig. \ref{fig3} for the case of $N=10$, we have two vortices in a
symmetric position related to the center of the sample and an external
ring with $8$ vortices. The vortices closer to the central ones have
bigger radius than those closer to the empty space between them. This is
easy to understand based on the competition between the different
interaction as we pointed out at the end of section II. When the external
filed is larger than $H_{s10}$, the repulsive interaction of the vortices
with the field that penetrates the surface is more important than the
vortex-vortex image interaction.  This interaction competes with the
repulsive interaction between the vortices which is more important between
the central vortices and the vortices of the outer ring that are closer
to them. This result agree with the result of Ternovskii {\it et al.}
\cite{ternovskii}. They showed for a semi-infinite plane, that the
vortex lattice is distorted near the surface. 

 As long as a
decade ago Yarmchuk {\it et al.} \cite{yarmchuk} determined experimentally
the vortex patterns up to $11$ vortices in superfluids and Campbell and
Ziff \cite{campbell} made a very extensive numerical study of these
patterns. Despite the interaction between vortices in a superconductor is
screened over on scale $\lambda$, and in superfluid it is logarithmic on
all scales, there is indeed a clear analogy between them. The patterns
found here are in excellent agreement with the experimental results found
in Ref. [\onlinecite{yarmchuk}]. The vortices does not show only similar
patterns but show the same tendency to vortex accumulation at the center
of the sample.  The same tendency has been observed in experiments with 
finite size superconducting samples
of different geometries \cite{brandt1,zeldov1} and this suggest that this
tendency is proper of finite samples independent of the particular
geometry. Our results are in good agreement also with the results of Ref.
[\onlinecite{campbell}], if we compare the ring structure but, two
differences we can point out. First, as they supposed predetermined ring
structures their vortex patterns does not show the detailed structure of
the outer rings as our more accurate calculations. Second, their vortex
patterns does not show the tendency of vortex accumulation at the center
of the sample. 

\section{SUMMARY}\label{sec6} 

In summary, by using London theory we have studied the size effects on a
long superconducting cylinder both in the mixed and Meissner state.  We
have determined numerically the ground state of the system and obtained
the vortex patterns, the induction and matching fields for $N=3$ up to
$N=18$ vortices. Our results show a clear tendency of vortex accumulation
in the center of the sample, delay in the penetration of the flux lines
and irregularities in the distribution of the vortices closer to the
surface, due to finite size effects. Those effects have been found before
in experiments with samples of different geometries showing, apparently,
that this is a consequence of the finite size of the samples and not of
the particular geometry.

\acknowledgments 
PAV thanks the Brazilian Agency Capes for partial
financial support. ES thanks the Brazilian Agencies CNPq and FAPESP for 
partial financial support. We thank Prof. Mauro M. Doria for very
useful discussions.

\newpage

\newpage
\begin{table}
\caption{Here $N_0$ represents a vortex at the center of the sample, $N_1$
and $N_2$ are the number of vortices on the first and second ring and
$R_1$ and $R_2$ are the respective radii of these rings. From 3 to 9 
vortices, all the
radii coincide up to the last digit shown in the table. Above 9 vortices,
the internal radii ($R_1$) coincide but, because of the fluctuations in the
value of the external radii, we have assumed $R_2$ as the mean value. The
seventh column is the energy associated to the terms that depend on the vortex positions and the last column represents the values of the critical fields for each
configuration of N vortices. The critical fields were determined with a
precision of $10^{-7}$ and the energies of $10^{-10}$, respectively, though these two quantities are quoted with seven decimals. Here we have used
the same parameters as in Fig. \protect\ref{fig2} and \protect\ref{fig3}. 
}\label{tab1}
\begin{tabular}{llllllll}
$N$ &$N_0$ &$N_1$ &$N_2$ &$R_1$ &$R_2$ & Energy &$H_{sN}$ \\
\hline
3&   & 3  &   & 3.2745 &      & 0.0348092  &3.7047593\\
4&   & 4  &   & 3.6834 &      & 0.0702077  &3.7157228\\
5&   & 5  &   & 4.0392 &      & 0.1283152  &3.7313060\\
6&  1& 5  &   & 4.5206 &      & 0.2092658  &3.7468671\\
7&  1& 6  &   & 4.6694 &      & 0.2931292  &3.7489583\\
8&  1& 7  &   & 4.833  &      & 0.4137502  &3.7722957\\
9&  1& 8  &   & 4.999  &      & 0.5784389  &3.8003082\\
10&  & 2  & 8 & 1.744  & 5.23 & 0.7557322  &3.8088703\\
11&  & 3  & 8 & 2.163  & 5.45 & 0.9599388  &3.8258413 \\
12&  & 3  & 9 & 2.146  & 5.53 & 1.1768955  &3.8335742 \\
13&  & 4  & 9 & 2.473  & 5.10 & 1.4435189  &3.8639324 \\
14&  & 4 & 10 & 2.455  & 5.77 & 1.7204882  &3.8702223 \\
15&  & 5 & 10 & 2.745  & 5.39 & 2.0593652  &3.9075828 \\
16&  & 5 & 11 & 2.724  & 5.97 & 2.4020796  &3.9106818 \\
17& 1& 5 & 11 & 3.178  & 5.60 & 2.7966109  &3.9419102 \\
18& 1& 6 & 11 & 3.314  & 5.71 & 3.1954771  &3.9465753 \\
\end{tabular}
\end{table}


\begin{references}
\bibitem[*]{santacruz} On leave from Departamento de
F\'{\i}sica, Universidade Estadual Paulista, Av. Engenheiro Luiz E. Coube
S/N, 17033-360 Bauru-SP, Brazil.
\bibitem{guyon} See for example E. Guyon and references therein in {\it Superconductivity}, Ed. P.R.
Wallace (Gordon and Breach - Science Publishers, 1969)
\bibitem{brandt1}E.\ H.\ Brandt, Rep.\ Prog.\ in Phys.\ {\bf 58}, 1465(1995) 
and references there in.
\bibitem{brandt2}E.\ H.\ Brandt, Phys.\ Rev.\ B {\bf 46}, 8628(1992).
\bibitem{brandt3}E.\ H.\ Brandt and M.\ Indembom, Phys.\ Rev.\ B {\bf 48}, 12893(1993).
\bibitem{shuster}Th.\ Schuster, H.\ Kuhn, E.\ H.\ Brandt and S.\ Klaum\"unzer, Phys.\ Rev. B {\bf 56},
3413(1997).
\bibitem{zeldov1}E.\ Zeldov, J.\ R.\ Clem, M.\ McElfresh, and M.\ Darvin, Phys.\ Rev.\ B {\bf 49}, 9802(1994).
\bibitem{zeldov2}E.\ Zeldov, A.\ I.\ Larkin, V.\ B.\ Geshkenbein, M.\ Konczykowsky, D.\ Majer, B.\ Khaykovich, V.\ M.\ Vinokur, and H.\ Shtrikman, Phys.\ Rev.\ Lett.\ {\bf 73}, 1428(1994).
\bibitem{exp1} B.\ P.\ Thrane,
C.\ Schlenker, J.\ Dumas, and R.\ Buder, Phys.\ Rev.\ B {\bf 54},
15518 (1996).
\bibitem{exp2} L.\ Civale, T.\ K.\ Worthington, and A.\ Gupta,
Phys.\ Rev.\ B {\bf 43}, 5425 (1991).
\bibitem{kaykovich}B.\ Kaykovich, E.\ Zeldon, M.\ Konczykowsky, D.\ Majer, 
A.\ I.\ Larkin, and J.\ R.\ Clem, Physica C,  {\bf 235-240}, 2757(1994).
\bibitem{jackson}J.\ D.\ Jackson, {\it Classical Eletrodynamics} 
(John Wiley \& Sons, New York, 1962). See Section 3.10, page 84.
\bibitem{buzdin}A.\ Buzdin, D.\ Feinberg Physica C, {\bf 256}, 303(1996).
\bibitem{brongersma}S.\ H.\ Brongersma, E.\ Verweij, N.\ J.\ Koeman, D.\ G.\ de 
Groot, R.\ Griessen and B.\ I.\ Ivlev Phys. Rev. Lett. {\bf 71}, 2319(1993).
\bibitem{numrec}W.\ H.\ Press, B.\ P.\ Flannery, S.\ A.\ Teukolsky,
and W.\ T.\ Vetterling, {\it Numerical Recipes} (Cambridge University
Press, Cambridge, England, 1992).
\bibitem{goffe}S.\ Kirkpatrick, C.\ D.\ Gelatt Jr, and M.\ P.\ Vecchi,
Science {\bf 220}, 671 (1993).
\bibitem{prudinokov}A.\ P.\ Prudinikov, Yu.\ A.\ Brychkov, and
O.\ I.\ Marichev, {\it Integrals and Series}, Gordon and Science
Publishers, Amsterdam (1986). See formula 5.910.
\bibitem{ternovskii} F.\ F.\ Ternovskii, L.\ N.\ Shekhata Sov. Phys. Jept {\bf 
35}, 1202(1972). 
\bibitem{yarmchuk} E.\ J.\ Yarmchuk, M.\ J.\ V.\ Gordon, and R.\ E.\ Packard,
Phys.\ Rev.\ Lett.\, {\bf 43}, 214(1979).
\bibitem{campbell}L.\ J.\ Campbell
and R.\ M.\ Ziff, Phys.\ Rev.\ B {\bf 20}, 1886 (1979).
\end{references}
\end{document}